\documentclass[twocolumn,aps,prl,amsmath,amssymb,twoside,showpacs,preprintnumbers]{revtex4}
\usepackage[T1]{fontenc}

\newcommand{\im}{\mathrm{i}}  
\newcommand{\xd}{\mathrm{d}}
\newcommand{\xD}{\mathcal{D}} 
\newcommand{\cS}{\mathcal{S}}
\newcommand{\cH}{\mathcal{H}}
\newcommand{\R}{\mathbb{R}}
\newcommand{\tens}{\otimes}

\newcommand{\be}{\begin{equation}}
\newcommand{\ee}{\end{equation}}

\begin{document}
\title{S-matrix at spatial infinity}
\author{Daniele Colosi}\email{colosi@matmor.unam.mx}
\author{Robert Oeckl}\email{robert@matmor.unam.mx}
\affiliation{Instituto de Matem\'aticas, UNAM, Campus Morelia,
C.P.~58190, Morelia, Michoac\'an, Mexico}
\date{16 Junly 2008 (v4)}
\preprint{UNAM-IM-MOR-2007-1}

\begin{abstract}
We provide a new method to construct the S-matrix in quantum field
theory. This method implements crossing symmetry manifestly by erasing
the a priori distinction between in- and out-states. It allows the
description of processes where the interaction weakens with distance
in space, but remains strong in the center at all times. It should
also be applicable to certain spacetimes where the conventional method
fails due to lack of temporal asymptotic states.
\end{abstract}

\maketitle

While the true state space of interacting quantum field theories usually
is not known, the S-matrix provides an excellent tool to describe
interactions using the state space of the corresponding free theory.
The S-matrix is defined as a map from an initial state
space of the free theory to a final one using the interaction picture
where free states are invariant 
under free time-evolution. Its elements are limits of
transition amplitudes between free states at early and at late times.
The interaction is only turned on at intermediate times and the limit
is determined by letting the initial and final times go to $-\infty$ and
$+\infty$ respectively.

This picture is adequate if the process to be described is such that the
interaction becomes negligible at very early and at very late times.
It is not adequate if the interaction remains important at all times.
An example of this would be a stationary bound state or, more extremely,
a stationary black hole. Another shortcoming of the conventional S-matrix
is that it cannot be defined for spacetimes that do not support temporal
asymptotic states, such as Anti-de-Sitter space.

We provide here a method to construct the S-matrix that is applicable to
processes where the interaction becomes negligible with distance from a center,
but may remain significant there at all times. This method relies on a notion
of \emph{spatial} asymptotic states that we will explain. It is also
applicable to certain
spacetimes that do not support temporal asymptotic states. For
example, it should allow to put the ``boundary S-matrix''
found for Anti-de-Sitter space \cite{Gid:smatrixadscft} on a solid
conceptual footing.
However, there are also situations where the new method does not apply,
but the conventional does. In particular, this is the case if space is
compact (e.g.\ de-Sitter space) or if spatial asymptotic states do not
exist for some other reason.

In this letter we present our method in Minkowski space and
show that it yields the same result there as the usual one when both
can be applied. This is the case when the
interaction can be neglected far away from the center both in
space and in time.
A conceptual advantage of our method over the conventional
one is that it
exposes crossing symmetry as a manifest feature of the S-matrix rather
than a derived one. 
This is due to the fact that in the present method there appears only
one state space in 
which in- and out-states are distinguished only by their quantum numbers.

Our method is based on the general boundary formulation of quantum
theory \cite{Oe:GBQFT} and its
application to quantum field theory \cite{Oe:timelike,Oe:KGtl}. The
mathematical
underpinning of this framework is a suitably adapted incarnation of
Segal's approach to
conformal field theory \cite{Seg:cftdef}. In short, this means we
associate states to
hypersurfaces in spacetime that are not restricted to be
spacelike. Also, for a region in spacetime
we associate an amplitude to a state that lives on its
boundary. Ordinary transition amplitudes
arise as special cases when the region in question is determined by a
time interval.
The mathematical formalism is supplemented by a physical
interpretation, which allows to extract
probabilities from generalized amplitudes \cite{Oe:GBQFT,Oe:probgbf}.

Two geometries are of relevance here: (a) the spacetime region
$[t_1,t_2]\times\R^3$ given by a
time interval $[t_1,t_2]$ extended over all of
space and (b) the spacetime region $\R\times B^3_R$, i.e., the ball of
radius $R$ in space extended over
all of time. We shall refer to the latter as the \emph{solid
  hypercylinder} and to its boundary
$\R\times S^2_R$ simply as the \emph{hypercylinder}. The first
geometry appears in the standard
transition amplitude between initial time $t_1$ and final time
$t_2$. State spaces, amplitudes and
probabilities associated with the second type of geometry where
introduced in \cite{Oe:KGtl} for free scalar quantum field theory.

We start by reviewing the ordinary approach to the S-matrix, deriving
it as a limit of transition
amplitudes between coherent states. This provides a blueprint for the
subsequent section, where
the asymptotic amplitude is derived that arises from coherent states
on hypercylinders of increasing
radius. Finally, the resulting amplitude and the conventionally derived
S-matrix are shown to coincide.

For simplicity, we limit ourselves to the massive real scalar
field. The generalization to
other types of fields should be straightforward. We use Schr\"odinger
(wave function) representations
for states \cite{Jac:schroedinger} and the Feynman path
integral for amplitudes. We mostly
follow conventions and notations of \cite{Oe:timelike,Oe:KGtl} and use
the techniques laid out there.

\section{Standard S-matrix}

The following derivation of the S-matrix is similar in spirit to the
method based on the holomorphic representation \cite{FaSl:gaugeqft}.
Standard
transition amplitudes take the form
\begin{multline}
 \langle \psi_2|U_{[t_1,t_2]}|
 \psi_1\rangle = N_{[t_1,t_2]}\\
 \cdot \int \xD\varphi_1\,\xD\varphi_2\, \psi_{1}(\varphi_1)
 \overline{\psi_{2}(\varphi_2)}
 \int_{\phi|_{t_i}=\varphi_i}\xD\phi\, e^{\im S(\phi)} ,
\label{eq:tampl1}
\end{multline}
where $U_{[t_1,t_2]}$ is the time-evolution operator from time $t_1$
to time $t_2$. $N_{[t_1,t_2]}$ is a normalization factor such that the
vacuum-to-vacuum amplitude equals one.
The outer integrals are over all field configurations
$\varphi_1$
and $\varphi_2$ in space. The inner integral is over all field
configurations $\phi$ in the spacetime region $[t_1,t_2]\times\R^3$
subject to the boundary conditions $\phi|_{t_1}=\varphi_1$ and
$\phi|_{t_2}=\varphi_2$.

We use the interaction picture to have a time independent description
of free states. It will be convenient to use coherent states
\cite{ItZu:qft}. A coherent state $\psi_\eta$ is
parametrized by a complex function $\eta$ on momentum space. At time
$t$ its wave function takes the form
\begin{multline}
\psi_{t,\eta} (\varphi) = N_{t,\eta}\\
 \cdot \exp \left( \int
 \frac{\xd^3 x\,\xd^3 k}{(2 \pi)^3} \, \eta(k) \, e^{-\im(E t-k x)} \,
 \varphi(x)\right)\, \psi_0(\varphi) ,
\end{multline}
where $\psi_0$ is the vacuum wave function and $N_{t,\eta}$ is a
normalization factor such that the state has unit norm.

Since we are in the interaction picture, the amplitude
(\ref{eq:tampl1}) for the free action is independent of initial and
final time and amounts to the
S-matrix of the free theory,
\begin{multline}
 \langle\psi_{\eta_2}|\cS_0|\psi_{\eta_1}\rangle=
 \exp \bigg( \int \frac{\xd^3k}{(2 \pi)^3 2 E} \,\\
 \left(  \eta_1(k)\, \overline{\eta_2(k)}  - \frac{1}{2} |\eta_1(k)|^2 -
 \frac{1}{2} |\eta_2(k)|^2\right) \bigg) .
\label{eq:sm0}
\end{multline}

We now modify the free theory by adding a source field $\mu$. That is, we
take the action
\be
 S_\mu(\phi)=S_0(\phi)+\int \xd^4 x\, \phi(x) \mu(x) ,
\label{eq:actionsrc}
\ee
where $S_0$ denotes the action of the free theory. The resulting
transition
amplitude again does not depend on initial and final time as long as
the source field is confined to the intermediate time
interval. Replacing the
free action in (\ref{eq:tampl1}) by (\ref{eq:actionsrc}) yields the
S-matrix of the theory with source,
\begin{multline}
 \langle\psi_{\eta_2}|\cS_\mu|\psi_{\eta_1}\rangle=
 \langle\psi_{\eta_2}|\cS_0|\psi_{\eta_1}\rangle\,
 \exp\left(\im\int \xd^4 x\,\mu(x)\hat{\eta}(x)\right)\\
 \cdot \exp\left(\frac{\im}{2}\int
 \xd^4 x\,\xd^4 x'\,
 \mu(x)G_F(x,x')\mu(x')\right),
\label{eq:smmu}
\end{multline}
where $G_F$ is the Feynman propagator normalized such that
$(\Box_x+M^2)G_F(x,x')=\delta^4(x-x')$.
$\hat{\eta}$ is the complex solution of the Klein-Gordon
equation given by
\begin{multline}
 \hat{\eta}(t,x)=\\
 \int \frac{\xd^3 k}{(2\pi)^3 2 E}
 \left(\eta_1(k) e^{-\im (E t- k x)}
 +\overline{\eta_2(k)} e^{\im(E t- k x)}\right) .
\label{eq:classcoh}
\end{multline}
We note that the initial and final coherent states determine
the positive and negative energy contributions to the
solution. Conversely, (\ref{eq:classcoh}) allows us to recover a pair
of coherent states from a complex solution.

This result combined with functional methods can be used to work out
the S-matrix for the general interacting theory. Consider the action
\begin{multline}
 S(\phi)=S_0(\phi)+\int \xd^4 x\, V(x,\phi(x)) =\\
 S_0(\phi)+\int\xd^4 x\,
 V\left(x,\frac{\partial}{\partial
 \mu(x)}\right) S_\mu(\phi)\bigg|_{\mu=0} ,
\label{eq:actionint}
\end{multline}
We assume at first that the interaction is cut off at early and at late
times, i.e., $V((t,x), a)=0$ if $t\le t_1$
or $t\ge t_2$. The transition amplitude
(\ref{eq:tampl1}) with (\ref{eq:actionint}) inserted then yields
\begin{multline}
 \langle\psi_{\eta_2}|\cS|\psi_{\eta_1}\rangle=
 \exp\left(\im\int\xd^4 x\,
 V\left(x,-\im\frac{\partial}{\partial
 \mu(x)}\right)\right)\\
 \langle\psi_{\eta_2}|\cS_\mu|\psi_{\eta_1}\rangle\bigg|_{\mu=0} .
\label{eq:smint}
\end{multline}
Again, there is no explicit dependence on $t_1$ or $t_2$ so we can
drop the cutoff and interpret the result as the S-matrix of the
interacting theory.

\section{S-matrix from timelike hypercylinders}

In this section we make heavy use of the results of \cite{Oe:KGtl}.
The amplitude associated with the solid 
hypercylinder of radius $R$ for a state $\psi$ takes the form,
\be
  \rho_R(\psi)
 = N_R \int \xD\varphi\, \psi_{R}(\varphi)
 \int_{\phi|_{R}=\varphi}\xD\phi\, e^{\im S(\phi)} .
 \label{eq:ampl}
\ee
where $N_R$ is a normalization factor, such that the amplitude of
the vacuum state equals one. The outer integral is
over field configurations $\varphi$ on the hypercylinder of radius
$R$. The inner integral is over field configurations $\phi$ in the
interior $\R\times B^3_R$ of the hypercylinder matching $\varphi$ on
the boundary.

The interaction picture is now defined by describing free states in a
radius-independent form. That is, we identify states
which are related by radial evolution. (To do this one needs the
amplitude associated to a hypertube \cite{Oe:KGtl}.) Again we use
coherent states. A coherent state $\psi_\xi$ in this setting may be
characterized
by a set of functions $\xi_{l,m}(E)$ that carry angular momentum
quantum numbers $l,m$ and depend on the energy $E$. The energy may be
positive or negative, but $\xi_{l,m}(E)=0$ if
$|E|<M$. The latter condition comes from the fact that the particle
spectrum on the hypercylinder is confined to non-negative values of
$p^2=E^2-M^2$ \cite{Oe:KGtl}. The wave function of the coherent state
at radius $R$ takes the form
\begin{multline}
\psi_{R,\xi} (\varphi) = N_{R,\xi} \exp\bigg(
 \int \xd t\, \xd \Omega\, \xd E \sum_{l,m}\\
  \xi_{l,m}(E)
 \frac{e^{\im E t}Y^{-m}_l(\Omega)}{2\pi h_l(p R)}
 \varphi(t,\Omega)\bigg)
\, \psi_{R,0}(\varphi) ,
\end{multline}
where $\psi_{R,0}$ is the vacuum wave function at radius $R$ and
$N_{R,\xi}$ is a normalization factor such that the state has unit
norm. Here, $Y^m_l$ denotes the spherical harmonic and $h_l$ the
spherical Bessel function of the third kind.
The inner product of coherent states is then
\begin{multline}
 \langle \psi_{R,\xi'},\psi_{R,\xi}\rangle=
 \exp\bigg(\int\xd E\sum_{l,m}\frac{p}{4\pi}
 \bigg(\xi_{l,m}(E)\, \overline{\xi_{l,m}'(E)}\\ 
 - \frac{1}{2} |\xi_{l,m}(E)|^2 - \frac{1}{2} |\xi_{l,m}'(E)|^2\bigg)
\bigg) ,
\end{multline}
Since we use the interaction picture, the amplitude of a state is
independent of the radius $R$,
\begin{multline}
 \cS_0(\psi_{\xi})=
 \rho_{R,0}(\psi_{R,\xi})=\exp\bigg(\int\xd E\sum_{l,m}\frac{p}{8\pi}\\
 \left(\xi_{l,m}(E)\xi_{l,-m}(-E)-|\xi_{l,m}(E)|^2\right)\bigg) .
 \label{eq:sm02}
\end{multline}

We turn to the theory with source field $\mu$ given by the action
(\ref{eq:actionsrc}). Working out the corresponding path integral
(\ref{eq:ampl}) when the source is confined to the interior of the
hypercylinder of radius $R$ yields,
\begin{multline}
 \cS_\mu(\psi_{\xi}) = \cS_0(\psi_{\xi})\,
 \exp\left(\im\int \xd^4 x\,\mu(x)\hat{\xi}(x)\right)\\
 \cdot\exp\left(\frac{\im}{2}\int
 \xd^4 x\,\xd^4 x'\,
 \mu(x)G_F(x,x')\mu(x')\right),
\label{eq:smmu2}
\end{multline}
where $G_F$ is the Feynman propagator. $\hat{\xi}$ is the complex solution of the Klein-Gordon equation
given by
\begin{multline}
 \hat{\xi}(t,r,\Omega)=\\
 \int \xd E\sum_{l,m}\frac{p}{2\pi}\xi_{l,m}(E)
 j_l(p r) e^{\im E t} Y^{-m}_l(\Omega) .
\label{eq:classcoh2}
\end{multline}
Here, $j_l$ denotes the spherical Bessel function of the first kind.
We note that this equation puts into correspondence coherent states
with complex solutions. Since (\ref{eq:smmu2}) does not depend on the
radius $R$, the restriction on $\mu$ to vanish outside of $R$ may
be lifted.

As before, we use functional methods to work out
the asymptotic amplitude for the general interacting theory. Take the
action (\ref{eq:actionint})
and assume at first that the interaction is cut off outside the radius
$R$, i.e., $V((t, x), a)=0$ if $|x|\ge R$. The amplitude  
(\ref{eq:ampl}) with (\ref{eq:actionint}) inserted then yields
\begin{multline}
 \cS(\psi_{\xi})=\\
  \exp\left(\im\int\xd^4 x\,
 V\left(x,-\im\frac{\partial}{\partial
 \mu(x)}\right)\right)
 \cS_\mu(\psi_\xi)\bigg|_{\mu=0} .
\label{eq:smint2}
\end{multline}
Again, there is no explicit dependence on $R$ so we can
drop the cutoff. Then (\ref{eq:smint2}) is the asymptotic
amplitude of the interacting theory.

\section{Equivalence of states and asymptotic amplitudes}

It is striking how much the expressions for the usual S-matrix and the
spatially asymptotic amplitude resemble each other.
It should be emphasized that this is a priori not at all obvious. In particular,
the fact that both in (\ref{eq:smmu}) and in
(\ref{eq:smmu2}) the same Feynman propagator appears is rather
non-trivial.

Let us explain this a little. The exponential term containing the Feynman
propagator in (\ref{eq:smmu}) arises in the form
\begin{equation}
\exp\left(\frac{\im}{2}\int
 \xd^4 x\, \mu(x)\alpha(x)\right).
\label{eq:alpha}
\end{equation}
where $\alpha$ is a complex solution of the inhomogeneous Klein-Gordon
equation $(\Box+M^2)\alpha=\mu$. $\alpha$ must satisfy boundary conditions
at early and at late time. Expanding $\alpha$ in terms of plane waves as
\begin{multline}
\alpha(t,x)=\\
\int \frac{\xd^3 k}{(2\pi)^3 2E} \left(\alpha^+(t,k) e^{-\im (E t - k x)}
+\alpha^-(t,k) e^{\im (E t - k x)}\right),
\end{multline}
the boundary conditions are $\alpha^+(t,k)=0$ for early times $t$ before the
source is switched on and $\alpha^-(t,k)=0$ for late times $t$ after the
source is switched off. These are just the Feynman boundary condition
leading to the substitution $\alpha(x)=\int \xd^4 x'\, G_F(x,x') \mu(x')$.

Similarly, in the hypercylinder setting the exponential containing the Feynman
propagator in (\ref{eq:smmu2}) also arises from an expression of the form
(\ref{eq:alpha}). Again $\alpha$
is a complex solution of the inhomogeneous Klein-Gordon equation. However, this
time the boundary conditions for $\alpha$ arise at large radius rather than at
large (early or late) time. Expanding $\alpha$ in terms of plane waves in time
and spherical harmonics in space,
\begin{multline}
 \alpha(t,r,\Omega)=
 \int \xd E\sum_{l,m}\left(\alpha^<_{l,m}(r,E)
 h_l(p r) e^{-\im E t} Y^{m}_l(\Omega)\right.\\
 \left. + \alpha^>_{l,m}(r,E) \overline{h}_l(p r) e^{\im E t} Y^{-m}_l(\Omega)\right),
\end{multline}
the boundary condition is $\alpha^>_{l,m}(r,E)=0$ for large radius $r$ outside
of the source field $\mu$. Surprisingly, this boundary condition turns out to be
equivalent to the Feynman boundary condition, yielding the same propagator.
We shall elaborate more on this elsewhere \cite{CoOe:smatrixgbf}.

If states and amplitudes associated with
hypercylinders yield as valid a description of physics as the usual
one we should be able to compare the descriptions and even conclude
their equivalence.
Indeed, consider a process (represented by an interaction) that is
bounded in space and in time. It should be possible to describe it either
via usual transition amplitudes for sufficiently early initial time
and late final time, or through amplitudes for a solid hypercylinder
of sufficiently large radius. What is
more, when we let the region where the process takes place grow
arbitrarily large (in space and in time) we should still get
equivalent results. This is indeed the case.

To compare the two settings we need a map
between the different boundary state spaces. In the standard setting
the boundary state space associated with the interval $[t_1,t_2]$ is
the tensor product $\cH_{t_1}\tens\cH_{t_2}^*$. (The dual at $t_2$ is
related to the fact that we should think of final states as bra-states
rather than ket-states.) We denote the state space associated with the
hypercylinder of radius $R$ by $\cH_R$. Thus, we are looking for an
isomorphism of Hilbert spaces $\cH_{t_1}\tens\cH_{t_2}^*\cong\cH_R$.
What is more, the relevant (asymptotic) amplitudes should be equal
under this isomorphism. Comparing (\ref{eq:smmu}) with
(\ref{eq:smmu2}) shows that this can be true only if the isomorphism
is given as follows: $\psi_{\eta_1}\tens\psi_{\eta_2}\cong\psi_\xi$ if
and only if
$\hat{\eta}=\hat{\xi}$. As we have remarked before,
(\ref{eq:classcoh}) and (\ref{eq:classcoh2}) really
establish bijective correspondences between classical solutions and
coherent states. Hence, the proposed isomorphism is really
bijective. Moreover, it is indeed an isomorphism (i.e., preserves the
inner product). One also easily checks that the free amplitudes
(\ref{eq:sm0}) and (\ref{eq:sm02}) are equal under the
isomorphism. The same follows for the amplitudes with source
(\ref{eq:smmu}) and (\ref{eq:smmu2}). Finally, it is then obvious
that the asymptotic amplitudes with general interaction (\ref{eq:smint})
and (\ref{eq:smint2}) also coincide.

Using the definition of particle states on the hypercylinder from
\cite{Oe:KGtl}, it is easy to verify that the isomorphism of state
spaces constructed above
indeed sends an in-state with $m$ particles and an out-state with $n$
particles to a states with $m+n$ particles on the hypercylinder. What
is more, the
latter have quantum numbers (the sign of the energy) that
identify them correctly as in- or out-going. However, the
meaning of in- or out-going is then with respect to the region defined
by the solid hypercylinder. This reinforces the notion that in- versus
out-going is not primarily a temporal property, but a spatiotemporal
one \cite{Oe:timelike,Oe:KGtl}. It indicates whether a particle goes
into or comes out of the (interaction) region of interest.

It is also clear how crossing symmetry in the hypercylinder case
is simply implicit in the fact that all particles are part of a single
state space. One could then view the isomorphism of state spaces as
yielding a derivation of crossing symmetry of the S-matrix
in the standard setup.

It remains to remark on the probabilities extracted from the
S-matrix. In \cite{Oe:KGtl} it was shown how the generalized
probability interpretation introduced in \cite{Oe:GBQFT} applies to
amplitudes associated with the solid hypercylinder. Using the methods
of \cite{Oe:KGtl} one can then show that the usual probability
interpretation of the S-matrix arises as a special case.

\acknowledgments

This work was supported in part by CONACyT grants 47857 and 49093.

\bibliography{stdrefs}
\bibliographystyle{amsordx}

\end{document}